\documentclass[aps,pra,twocolumn,superscriptaddress,showpacs]{revtex4-1}
\usepackage{graphicx}
\usepackage{amsmath}
\usepackage{nicefrac}
\usepackage{upgreek}
\begin{document}


\newcommand{\barb}{\overline{b}}
\newcommand{\barbh}{\overline{b}^{(0)}}
\newcommand{\barpsi}{\overline{\psi}}
\newcommand{\barphi}{\overline{\phi}}
\newcommand{\barbeta}{\overline{\beta}}
\newcommand{\barGamma}{\overline{\Gamma}}
\newcommand{\pone }{{P^{(1)}}}
\newcommand{\ptwo}{{P^{(2)}}}

\preprint{pra}

\title{Quantum frequency conversion and strong coupling of photonic modes using four-wave mixing in integrated microresonators}

\author{Z. Vernon}
\email{zachary.vernon@utoronto.ca}
\affiliation{Department of Physics, University of Toronto, 60 St. George Street, Toronto, Ontario, Canada, M5S 1A7}
\author{M. Liscidini}
\affiliation{Department of Physics, University of Pavia, Via Bassi 6, Pavia, Italy}
\author{J.E. Sipe}
\affiliation{Department of Physics, University of Toronto, 60 St. George Street, Toronto, Ontario, Canada, M5S 1A7}

\date{\today}

\begin{abstract}
Single photon-level quantum frequency conversion has recently been demonstrated using silicon nitride microring resonators. The resonance enhancement offered by such systems enables high-efficiency translation of quantum states of light across wide frequency ranges at sub-watt pump powers. Using a quantum-mechanical Hamiltonian formalism, we present a detailed theoretical analysis of the conversion dynamics in these systems, and show that they are capable of converting single- and multi-photon quantum states. Analytic formulas for the conversion efficiency, spectral conversion probability density, and pump power requirements are derived which are in good agreement with previous theoretical and experimental results. We show that with only modest improvement to the state of the art, efficiencies exceeding 95\% are achievable using less than 100 mW of pump power. At the critical driving strength that yields maximum conversion efficiency, the spectral conversion probability density is shown to exhibit a flat-topped peak, indicating a range of insensitivity to the spectrum of a single photon input. Two alternate theoretical approaches are presented to study the conversion dynamics: a dressed mode approach that yields a better intuitive picture of the conversion process, and a study of the temporal dynamics of the participating modes in the resonator, which uncovers a regime of Rabi-like coherent oscillations of single photons between two different frequency modes. This oscillatory regime arises from the strong coupling of distinct frequency modes mediated by coherent pumps.
\end{abstract}

\pacs{[PACS Nos. here]}
\maketitle

\section{Introduction}\label{sec:intro}
Reliable control over quantum states of light is an important objective of the optics community. The ability to deterministically manipulate the degrees of freedom in each photon of an optical state is of paramount importance for quantum optical technologies. Strategies for achieving such control are at the core of efforts to advance optical quantum information processing and computing. 

One crucial attribute of a photon is its frequency. This degree of freedom can be used to encode information \cite{Clemmen2016}, can serve as an entanglement resource \cite{Azzini2012,Takesue2004,Ramelow2015}, or may simply be chosen through design considerations of source, transmission, or detection: selecting a convenient frequency range for a specific experiment may depend on the technology that already exists. For example, it is often desirable to work in the telecommunications band with frequencies near 193 THz -- corresponding to a wavelength of 1550 nm -- where many commercially available devices efficiently and accurately operate. However, for quantum applications that require single photon detection, working in this band necessitates the use of expensive cryogenically cooled superconducting single photon detectors. It is therefore desirable to construct a simple and inexpensive device that translates quantum states of light from the telecommunications band to the wavelength range of 600-800 nm, where inexpensive room temperature-operated silicon avalanche photodetectors work efficiently \cite{Hadfield2009,Albota2004}.

The process in which single photons (or, more generally, quantum states of light) are translated in frequency is termed quantum frequency conversion (QFC). Conventional high-efficiency implementations of QFC rely on bulk nonlinear optical elements \cite{Albota2004,Huang1992} or long fibres \cite{Mcguiness2010,Clark2013}, and typically require one or several watts of pump power to attain high conversion efficiencies. QFC at mW-level powers has been achieved using integrated nanowires \citep{Agha2013}, but obtaining high efficiencies using such short interaction media remains challenging. A compact, integrated chip-based device that operates at sub-watt pump powers and attains near-unit conversion efficiency with low noise would therefore represent an important advance in QFC technology. 

These needs can be met by integrated resonant microstructures. By taking advantage of the resonant field enhancement offered by such systems, the input pump power needed to achieve high conversion efficiency can be drastically lowered, as proposed by Huang \emph{et al.} \cite{Huang2013}. This idea came to fruition in a recent  experiment carried out by Li \emph{et al.} \cite{Li2016}, wherein a silicon nitride microring resonator was employed to convert a weak, single photon-level input signal near 1550 nm to near 980 nm using less than 60 mW of pump power with a conversion efficiency exceeding 60\%. Diamond microresonators have also been proposed as a system for converting single photons produced by silicon-vacancy colour centres to telecommunications bands \cite{Lin2015}.

In this paper we present a theoretical study of the dynamics of quantum frequency conversion in microresonators. While we focus on single-photon input states in microring resonators, we show that microresonator-based QFC devices are capable of translating multi-photon quantum states across large frequency ranges. As is required due to the large span of wavelengths involved in QFC, our model includes the effects of different coupling conditions, quality factors, and loss rates across different modes. We develop our formalism for QFC schemes that exploit four-wave mixing arising from the third-order nonlinear optical response, but our techniques can easily be extended to treat media with second-order nonlinearities, such as aluminum nitride \cite{Xiong2012, Guo2015}. Our results indicate that wideband QFC, with near-unit efficiency using under 100 mW of pump power, is possible in silicon nitride microring resonators close to the current state of the art. By studying the strongly driven regime in this system, it is also possible to identify effects that arise from the strong coupling of different frequency modes. This enables the exploration of phenomena in an all-photonic platform that are usually only observed in driven fermionic systems, such as isolated atoms or quantum dots coupled to optical resonators, and quantum wells.

In Sec. \ref{sec:basic_eqns} we begin with the full Hamiltonian that describes the QFC process, including all linear and nonlinear terms, as well as those which describe the effects of scattering loss. In Sec. \ref{sec:struct_design} we then  discuss the subtleties of device design, including dispersion considerations and unwanted effects that lead to noise in the device output. In Sec. \ref{sec:conv_dynamics} we use a frequency-domain approach to calculate the spectral conversion probability density, conversion probability, and power requirements for QFC, comparing our predictions to the experimental results of Li \emph{et al.} We then develop two alternate approaches to study the conversion dynamics: in Sec. \ref{sec:dressed_modes} we construct a dressed mode picture that yields a more intuitive explanation for the qualitative behaviour of the QFC process, and in Sec. \ref{sec:temporal_dynamics} we study the time evolution of the intraring photon number expectation values, confirming the regime of Rabi-like oscillations as a single photon input oscillates between different frequency modes. Our results are summarized and areas for future work are discussed in Sec. \ref{sec:conclusion}.

\section{System Hamiltonian}\label{sec:basic_eqns}
In this section we lay out the Hamiltonian that describes the microring-channel system. We begin with the linear terms, for which the essential points are summarized; for a detailed discussion of these the reader is referred to our earlier work \cite{Vernon2015,Vernon2015b} and other treatments of microresonator quantum optics \cite{Camacho2012,Matsko2005,Chembo2010,Chembo2016}. We then discuss in detail the nonlinear interaction that yields the desired frequency conversion process.
\subsection{Linear Hamiltonian}\label{subsec:Linear_hamiltonian}
\begin{figure}
\includegraphics[width=1.0\columnwidth]{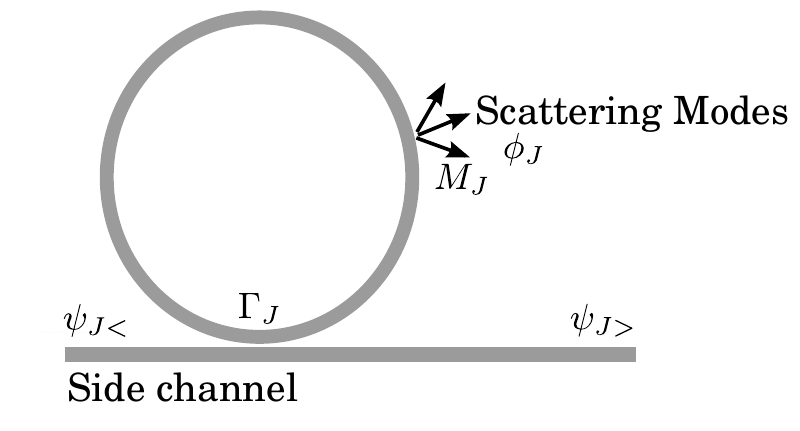}
\caption{Schematic of ring-channel structure for quantum frequency conversion. Input fields $\psi_{J<}$ interact with the modes $J$ in the ring and exit the system into the outgoing fields $\psi_{J>}$. The coupling rate associated with the ring-channel coupling is $\Gamma_J$; photons in the ring can also be lost to scattering fields $\phi_J$, with associated rates $M_J$. The FWHM linewidth $\Delta f_J^\mathrm{FWHM}$ of ring mode $J$ is then $\Delta f_J^\mathrm{FWHM}=(\Gamma_J + M_J)/\pi=\barGamma_J/\pi$.}
\label{fig:ring_schematic}
\end{figure}

We consider a microring resonator with radius $R$ side-coupled to a single channel waveguide as illustrated in Fig. \ref{fig:ring_schematic}. The full system Hamiltonian $H$ is divided into constituent components according to \cite{Vernon2015}
\begin{eqnarray}\label{eqn:full_H}
H = H_\mathrm{channel}+ H_\mathrm{ring} + H_\mathrm{coupling} + H_\mathrm{loss},
\end{eqnarray}
where $H_\mathrm{channel}$ describes the fields propagating in the side channel, $H_\mathrm{ring}$ the resonator modes, and $H_\mathrm{coupling}$ their coupling to the channel fields. Finally, scattering modes into which ring photons can be lost, as well as the couplings of those modes to the ring modes, are described by $H_\mathrm{loss}$. The ring accommodates a comb of modes $J$ with circular frequencies $\omega_J=2\pi f_J$ and wavenumbers $k_J$ that satisfy the resonance condition
\begin{eqnarray}\label{eqn:quantization}
k_J=\frac{2\pi m_J}{2\pi R}=\frac{m_J}{R},
\end{eqnarray}
where $m_J$ is a positive integer corresponding to the order of the mode $J$. We assume the radius $R$ is sufficiently small that the free spectral range between neighbouring modes greatly exceeds each resonance linewidth within the entire mode spectrum; that is, we are in the high finesse regime at all frequencies of interest. Each mode $J$ is then represented by a corresponding annihilation operator $b_J$, giving rise to a ring Hamiltonian of the form
\begin{eqnarray}\label{eqn:ring_H}
H_\mathrm{ring} = \sum_J \hbar\omega_J b_J^\dagger b_J + H_\mathrm{NL},
\end{eqnarray}
where $H_\mathrm{NL}$ contains all the nonlinear interaction terms between the ring modes. Using the ring resonances as reference frequencies, we can write the channel Hamiltonian as a sum over terms involving field operators $\psi_J(z)$ that only contain modes with frequencies near $\omega_J$. These fields then obey to very good approximation the usual commutation relations
\begin{eqnarray}\label{eqn:field_commutator}
\left[\psi_J(z),\psi^\dagger_{J'}(z')\right]&=&\delta_{JJ'}\delta(z-z'), \nonumber \\
\left[\psi_J(z),\psi_{J'}(z')\right]&=&0,
\end{eqnarray}
which allows us to write $H_\mathrm{channel}$ as
\begin{eqnarray}\label{eqn:channel_H}
\lefteqn{H_\mathrm{channel} =} \\
& & \sum_J\Bigg( \hbar\omega_J\int dz \psi_J^\dagger(z) \psi_J(z) \nonumber \\
&+& \frac{i\hbar v_J}{2}\int dz \left[ \frac{d\psi_J^\dagger(z)}{dz}\psi_J(z) - \mathrm{H.c.} \right] \Bigg) \nonumber,
\end{eqnarray}
where $v_J$ is the group velocity for channel modes with frequencies near $\omega_J$. Our model thus accounts for material and modal dispersion between these frequencies, but assumes that the group velocity does not vary significantly within the linewidth of an individual ring resonance. 

The Hamiltonian $H_\mathrm{coupling}$ describing the coupling between the ring and channel can be written as \cite{HamiltonianNote}
\begin{eqnarray}\label{eqn:coupling_H}
H_\mathrm{coupling}=\sum_J\left(\hbar\gamma_J^* b_J^\dagger\psi_J(0) + \mathrm{H.c.}\right),
\end{eqnarray}
in which we have approximated the ring-channel coupling as occurring at a single point $z=0$. The coefficients $\gamma_J$ determine the coupling strength between the channel fields and ring modes, and can be controlled by fabricating structures with different ring-channel coupling gaps, and by modifying the effective length over which the evanescent fields from the ring and channel overlap. These coefficients are related to the extrinsic quality factors $Q^\mathrm{ext}_J$ of the ring modes via 
\begin{eqnarray}\label{eqn:extrinsic_Q}
Q^\mathrm{ext}_J = \frac{\omega_J}{2\Gamma_J},
\end{eqnarray}
where $\Gamma_J=|\gamma_J|^2/2v_J$ is the rate associated with the ring-channel coupling. These extrinsic quality factors differ from the full quality factors $Q_J$, which incorporate both $Q^\mathrm{ext}_J$ as well as the \emph{intrinsic} quality factors $Q^{\mathrm{int}}_J$ that arise from the effects of scattering losses in the ring  \cite{CouplingRateNote}.
A convenient way to model such losses in this system \cite{Vernon2015} is to introduce a fictitious ``phantom channel", identical to the physical channel, accommodating fields $\phi_J(z)$ with group velocities $u_J$ that couple to the ring in exactly the same manner as represented in $H_\mathrm{coupling}$, but with coupling coefficients $\mu_J$ in place of $\gamma_J$. The intrinsic quality factor is then given by
\begin{eqnarray}\label{eqn:intrinsic_Q}
Q^\mathrm{int}_J=\frac{\omega_J}{2M_J},
\end{eqnarray}
where $M_J=|\mu_J|^2/2u_J$ is the coupling rate associated with scattering. The full, loaded quality factor $Q_J$ of each mode then obeys
\begin{eqnarray}\label{eqn:full_Q}
\frac{1}{Q_J} = \frac{1}{Q^\mathrm{ext}_J} + \frac{1}{Q^\mathrm{int}_J},
\end{eqnarray}
which gives $Q_J=\omega_J/(2\barGamma_J)$, with $\barGamma_J=\Gamma_J + M_J$ the total damping rate of mode $J$; the FWHM linewidth $\Delta f_J^\mathrm{FWHM}$ of mode $J$ is then simply $\barGamma_J/\pi$.

When studying the quantum statistics of photons generated in the ring it is crucial to distinguish between the extrinsic and intrinsic quality factors, as the relative magnitudes of the associated coupling rates have a drastic impact on single-photon detection probabilities \cite{Vernon2015,Vernon2016}. Indeed, as will become apparent in Sec. \ref{sec:conv_dynamics}, for the purposes of single photon frequency conversion it is necessary to construct a strongly over-coupled microring structure, in which $Q^\mathrm{int}_J \gg Q^\mathrm{ext}_J$ for all relevant modes $J$, ensuring that single photons in the microring predominantly couple out to the side channel rather than are lost to scattering.

\subsection{Nonlinear Interaction}\label{subsec:NL_sector}
\begin{figure}
\includegraphics[width=1.0\columnwidth]{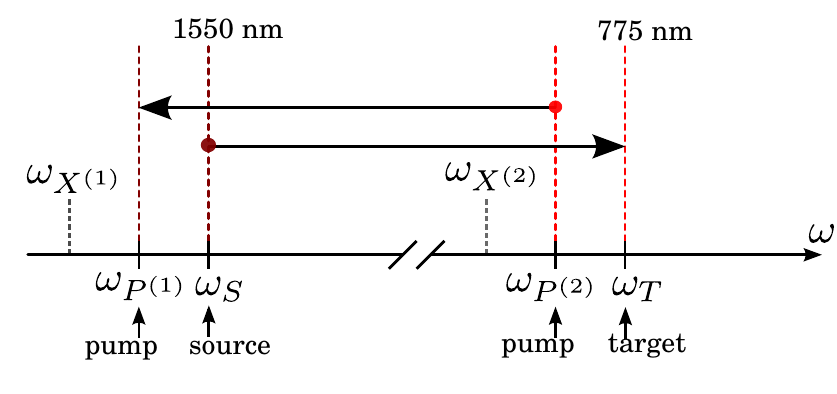}
\caption{Schematic of frequency structure for quantum frequency conversion. A single photon in the source mode centred at frequency $\omega_S$ is transferred to the target mode at $\omega_T$. This process is accompanied by a photon from the strongly pumped mode $\ptwo$ at $\omega_\ptwo$ being transferred to the strongly pumped mode $\pone $ at $\omega_\pone $. Also illustrated are frequencies $\omega_{X^{(1,2)}}$, which must be at least several linewidths away from any ring resonance to prevent spurious photons being generated in the source and target modes (see Sec. \ref{sec:struct_design}). Per Eq. (\ref{eqn:energy_conservation}), the frequency separation between the two pumped modes must equal that between the source and target for the process to conserve energy. As expressed in Eq. (\ref{eqn:order_matching}), phase matching requires mode $\pone $ to be separated from the source by the same number of mode orders as the target is from $\ptwo$. Note that the wavelengths 1550 nm and 775 nm are labelled here for illustrative purposes only; the conversion scheme does not sensitively depend on the specific choice of wavelengths.}\label{fig:mode_scheme} 
\end{figure}

Here we consider a particular conversion scheme that involves four different modes. The \emph{source} photon centred at the resonant frequency $\omega_S$ is injected into the ring through the channel, and is up-converted to the \emph{target} photon centred at resonant frequency $\omega_T$. The up-conversion results from the third-order nonlinear interaction between source and target photons mediated by two additional strong coherent beams at resonant frequencies $\omega_{P^{(1)}}$ and $\omega_{P^{(2)}}$. These are illustrated in Fig. \ref{fig:mode_scheme}. The relevant term in the nonlinear part of the Hamiltonian is \cite{HamiltonianNote}
\begin{equation}\label{eqn:nonlinear_H}
H_\mathrm{NL} = -\hbar\Lambda\left(b_Sb_\ptwo b_T^\dagger b_\pone ^\dagger + \mathrm{H.c.}\right),
\end{equation}
where $\Lambda$ is the nonlinear coupling strength parameter; this can be estimated as $\Lambda \approx 2\hbar\overline{\omega}^2 c n_2/(\overline{n}^2 V_\mathrm{ring})$, where $\overline{\omega}^2=\sqrt{\omega_S\omega_\pone \omega_T\omega_\ptwo }$, $\overline{n}^2=\sqrt{n(\omega_S)n(\omega_{P^{(1)}})n(\omega_T)n(\omega_{P^{(2)}})}$ with $n(\omega)$ the linear refractive index of the ring material at $\omega$, $n_2$ the nonlinear refractive index of the ring material, and $V_\mathrm{ring}$ the volume of the ring mode \cite{Andersen2015,Vernon2015b}. This estimate for $\Lambda$ assumes near-perfect phase matching of the nonlinear interaction, which requires the wavenumbers of these modes to satisfy 
\begin{eqnarray}\label{eqn:phase_matching}
k_\ptwo  - k_\pone  = k_T - k_S,
\end{eqnarray}
meaning the corresponding mode orders must obey
\begin{eqnarray}\label{eqn:order_matching}
m_\ptwo  - m_\pone  = m_T - m_S.
\end{eqnarray}

For this process to conserve energy, the frequency separation of the pumped modes must equal that between the source and target;
\begin{eqnarray}\label{eqn:energy_conservation}
\omega_\ptwo  - \omega_\pone  = \omega_T - \omega_S.
\end{eqnarray}
The process described by (\ref{eqn:nonlinear_H}) physically corresponds to a photon in $\ptwo$ being transferred to $\pone $, while the source photon is simultaneously transferred to the target mode. The bright, coherent nature of the energy in the pumped modes strongly couples the source and target modes, mediating an effective beamsplitter-like interaction that transfers photons from the source to the target. This interaction is sometimes referred to as Bragg scattering four-wave mixing (BS-FWM) \cite{Marhic1996,Li2016}.

 Note that many additional terms besides (\ref{eqn:nonlinear_H}) appear in $H_\mathrm{NL}$, including terms that describe self-phase modulation, cross-phase modulation, and various  four-wave mixing processes which transfer photons between other ring modes. However, we show in the Appendix that the effects of self- and cross-phase modulation can be easily cancelled by suitably adjusting the frequencies and intensities of the pump input beams. The frequencies $\omega_J$ in this work are thus understood to include the effects of self- and cross-phase modulation, which usually lead to modest frequency offsets at the powers considered. Furthermore, as detailed in the following section, by taking advantage of the significant dispersion it is possible to find a specific pair of pump modes that will suppress any other competing four-wave mixing processes that influence the source and target modes. Keeping in mind these considerations, it is justified to study the desired interaction (\ref{eqn:nonlinear_H}) in isolation, neglecting the other terms that appear in the full nonlinear Hamiltonian.
 
We choose to study quantum frequency conversion using this particular nonlinear interaction due to its ability to translate single- and multi-photon states across large frequency ranges using modest input powers and with noise limited only by technical rather than fundamental considerations. As will become clear in the following section, this interaction also enables the strong coupling of photonic modes with very different frequencies.

\section{Structure Design}\label{sec:struct_design}
While in principle any two modes satisfying (\ref{eqn:energy_conservation}) and (\ref{eqn:phase_matching}) can be chosen for the pumps, in practice they should be selected to lie as far as possible in frequency from the source and target. This criterion is important for minimizing the contamination of the source and target modes with spurious photons generated via spontaneous Raman scattering (SRS) of photons from the pumped modes. By selecting $\pone $ to have lower frequency than the source, and $\ptwo$ to have lower frequency than the target, the Stokes contribution to noise in the source and target modes can be eliminated, leaving only the anti-Stokes contribution, which is minimized by separating the pumped modes from the source and target. The $\pone $ mode must be separated from the source by exactly the same number of mode orders as $\ptwo$ is from the target to satisfy the phase matching constraint (\ref{eqn:phase_matching}); simultaneously satisfying energy conservation (\ref{eqn:energy_conservation}) therefore requires the parameters of the waveguide structure out of which the microring is formed to be designed such that the free spectral range near the source mode equals that near the target. 

An additional restriction on the choice of pumped modes arises from undesired four-wave mixing processes. Apart from being unaffected by noise from SRS, the source and target modes must also remain uncontaminated by any $\chi_{(3)}$ process that results in a photon emitted into those modes, other than the desired interaction (\ref{eqn:nonlinear_H}). Several such possible processes can be identified:
\begin{enumerate}
	\item Two $\pone $ photons may produce a pair of photons, one of which may be at the source frequency ($2\omega_\pone \rightarrow \omega_S + \omega_{X^{(1)}}$ where $X^{(1)}$ is a possible unwanted ring mode).
	\item Two $\ptwo$ photons may produce a pair of photons, one of which may be at the target frequency ($2\omega_\ptwo \rightarrow \omega_T + \omega_{X^{(2)}}$, where $X^{(2)}$ is a possible unwanted ring mode).
\end{enumerate}
Were the free spectral range of the ring resonator modes uniform over its entire span, eliminating these parasitic effects would be impossible: there would always exist modes at the undesired frequencies $\omega_{X^{(1,2)}}$. However, the modal and material dispersion can impose a significant variation in the mode spacing in different regions of the mode comb. The local free spectral range between neighbouring modes $J$ and $J'$ is given by
\begin{eqnarray}\label{eqn:FSR_calculation}
\lefteqn{\Delta f_J^{\mathrm{FSR}}} \nonumber \\
&=&\frac{1}{2\pi}(\omega_J(k_{J'}) - \omega_J(k_{J})) \nonumber \\
&\approx& \frac{1}{2\pi}\left[ \omega_J(m_J/R) + \frac{d\Delta}{dk}\bigg\vert_{k=k_J}\left(\frac{m_J+1}{R}-\frac{m_J}{R}\right)\right] \nonumber \\
 & &- \frac{1}{2\pi}\omega_J(m_J/R) \nonumber \\
&=& \frac{v_J^{\mathrm{ring}}}{2\pi R},
\end{eqnarray}
where $v_J^\mathrm{ring}$ is the group velocity in the ring associated with mode $J$. The free spectral range is thus proportional to the local (in frequency space) group velocity. While the microring must be engineered to have equal group velocities near the source and target modes, it is important to ensure that the group velocities near those modes do vary by an amount sufficient to ensure the absence at the undesired frequencies $\omega_{X^{(1,2)}}$. This effect is evident in the experimental data reported by Li \emph{et al.} \cite{Li2016}, wherein output photons were observed at sidebands situated symmetrically about the pump modes and displaced in frequency opposite the source and target modes. The amount of generated power in these sidebands was observed to decrease as the source photon frequency was translated farther from the corresponding pump; this can be attributed to the growing frequency mismatch that arises from dispersion in the ring.

 With a properly designed structure, either simulated or from an experimentally characterized microring, it is possible to find a pair of pump modes which satisfy the desired energy-conserving relation (\ref{eqn:energy_conservation}), and are separated from the source and target modes by an equal number of mode orders, but for which no modes at the undesired frequencies $\omega_{X^{(1,2)}}$ exist \cite{Li2016}. Of course, for realistic mode structures these conditions cannot be perfectly satisfied; however, it suffices that (\ref{eqn:energy_conservation}) holds to a precision within the linewidth of the resonator modes, and that no modes exist at frequencies within several linewidths of the undesired frequencies $\omega_{X^{(1,2)}}$. 

\section{Conversion Dynamics}\label{sec:conv_dynamics}
To calculate the properties of the QFC device, such as the probability of a source photon being successfully transferred to the target mode, we solve the relevant Heisenberg equations of motion for the slowly-varying ring operators $\barb_J(t)=b_J(t)e^{i\omega_Jt}$, treating the pumps classically while retaining the quantum-mechanical nature of the source and target modes. By introducing incoming and outgoing slowly-varying channel field operators $\barpsi_{J<}(z,t)$ and $\barpsi_{J>}(z,t)$ that respectively correspond to the channel fields before and after the ring-channel coupling point \cite{Vernon2015}, the fields immediately to the right of the coupling point can be calculated via
\begin{eqnarray}\label{eqn:channel_transformation}
\barpsi_{J>}(0,t) = \barpsi_{J<}(0,t) - \frac{i\gamma_J}{v_J}\barb_J(t).
\end{eqnarray}
The source and target mode operators in the ring then satisfy a simple set of coupled ordinary differential equations,
\begin{subequations}\label{eqn:S_T_ring_eqns}
\begin{eqnarray}
\lefteqn{\left(\frac{d}{dt} + \barGamma_S \right)\barb_S(t)=} \\
& &-i\gamma_S^*\barpsi_{S<}(0,t) -i\mu_S^*\barphi_{S<}(0,t) + i\Lambda\barbeta_\ptwo ^*\barbeta_\pone  \barb_T(t), \nonumber \\
\lefteqn{\left(\frac{d}{dt} + \barGamma_T \right)\barb_T(t)=} \\
& &-i\gamma_T^*\barpsi_{T<}(0,t) -i\mu_T^*\barphi_{T<}(0,t) + i\Lambda\barbeta_\ptwo \barbeta_\pone ^*\barb_S(t) \nonumber,
\end{eqnarray}
\end{subequations}
where $\barbeta_\ptwo $ and $\barbeta_\pone $ are the amplitudes of the pumped modes in the ring. We assume these pumped modes are driven by classical, resonant cw beams, and have reached a constant steady-state amplitude in the ring; these amplitudes are then given by
\begin{eqnarray}\label{eqn:pump_amplitudes}
\barbeta_J=\frac{-i\gamma_J^*e^{i\xi_J}}{\barGamma_J}\sqrt{\frac{P^\mathrm{in}_{J}}{\hbar\omega_J v_J}},
\end{eqnarray}
where $P^\mathrm{in}_{J}$ is the input power in mode $J$ (either $P^{(1)}$ or $P^{(2)}$) and $\xi_J$ is the (constant) phase of the associated beam in the channel.

The equations (\ref{eqn:S_T_ring_eqns}) can easily be solved in the frequency domain: defining for any slowly-varying operator $\overline{\mathcal{O}}(t)$ the Fourier amplitude $\hat{\mathcal{O}}(\Delta)$,
\begin{eqnarray}
\hat{\mathcal{O}}(\Delta)=\int \frac{dt}{\sqrt{2\pi}} \overline{\mathcal{O}}(t)e^{i\Delta t},
\end{eqnarray}
we obtain for the ring operators
\begin{subequations}\label{eqn:S_T_freqdomain_eqns}
\begin{eqnarray}
\lefteqn{\left(-i\Delta + \barGamma_S\right)\hat{b}_S(\Delta)=} \\
& &-i\gamma_S^*\hat{a}_S(\Delta) - i\mu_S^*\hat{d}_S(\Delta) + ig\hat{b}_T(\Delta), \nonumber \\
\lefteqn{\left(-i\Delta + \barGamma_T\right)\hat{b}_T(\Delta)=} \\
& &-i\gamma_T^*\hat{a}_T(\Delta) - i\mu_T^*\hat{d}_T(\Delta) + ig^*\hat{b}_S(\Delta), \nonumber
\end{eqnarray}
\end{subequations}
where $g=\Lambda\barbeta_\ptwo ^*\barbeta_\pone $ is the source-target coupling parameter. The $\hat{a}_J(\Delta)$ are the annihilation operators for modes with frequencies $\omega_J+\Delta$ in the incoming channel field $J$, and $\hat{d}_J(\Delta)$ are similar annihilation operators for the phantom channel fields; since the fast optical frequencies have been removed from the barred operators, \emph{the variable $\Delta$ now represents a frequency offset from the relevant ring mode reference frequency.} While the channel operators satisfy the commutation relations
\begin{eqnarray}\label{eqn:freq_commutator}
\left[\hat{a}_J(\Delta),\hat{a}_{J'}(\Delta')\right] &=& 0, \nonumber \\
\left[\hat{a}_J(\Delta),\hat{a}_{J'}^\dagger(\Delta')\right] &=& v_J^{-1}\delta_{JJ'}\delta(\Delta-\Delta'),
\end{eqnarray}
and similar for $\hat{d}_J(\Delta)$, the frequency-domain ring operators $\hat{b}_J(\Delta)$ do \emph{not} satisfy any such simple commutation relations.

We are primarily interested in the contribution to $\hat{b}_T(\Delta)$ from $\hat{a}_S(\Delta)$; solving this system of algebraic equations, we obtain for the target
\begin{eqnarray}\label{eqn:b_T_soln}
\hat{b}_T(\Delta)=\frac{\gamma_S^*g^*}{(-i\Delta + \barGamma_T)(-i\Delta + \barGamma_S)+|g|^2}\hat{a}_S(\Delta),
\end{eqnarray}
in which we have neglected all terms involving the other channel fields, since those terms will not contribute to the quantities of interest in this work. Using the channel input-output relation (\ref{eqn:channel_transformation}), for the outgoing target mode annihilation operators in the channel $\hat{c}_T(\Delta)$ (keeping only the term involving $\hat{a}_S(\Delta)$) we obtain
\begin{eqnarray}\label{eqn:c_T_expression}
\hat{c}_T(\Delta) = \frac{i\gamma_T\gamma_S^*g^*/v_T}{(-i\Delta+\barGamma_T)(-i\Delta+\barGamma_S)+|g|^2}\hat{a}_S(\Delta).
\end{eqnarray}

The omission of terms involving other channel field amplitudes, such as $\hat{a}_T(\Delta)$, is justified provided we restrict ourselves to using (\ref{eqn:c_T_expression}) only to calculate physical quantities relating to the outgoing target field for inputs that are confined to frequencies close to the source mode. This relation enables the calculation of properties of the outgoing target field for \emph{any} quantum state input of the source field, not merely single photon states: QFC can be used to convert N-photon Fock states, squeezed states, or other multi-photon quantum optical inputs. For an arbitrary input state $\vert \Phi_S\rangle$ with frequency support confined to a bandwidth near the the source mode frequency, the expectation value of an arbitrary normal-ordered operator product of the form $\mathcal{O}(\nu_1,...,\nu_N)=\prod_{j=1}^{M}\hat{c}_T^\dagger(\nu_j)\prod_{k=M+1}^N\hat{c}_T(\nu_k)$ is given by
\begin{eqnarray}\label{eqn:arbitrary_expect_value}
\lefteqn{\langle\mathcal{O}(\nu_1,...,\nu_N)\rangle} \nonumber \\
&=&\langle \Phi_S \vert \prod_{j=1}^{M}\hat{c}_T^\dagger(\nu_j)\prod_{k=M+1}^N\hat{c}_T(\nu_k) \vert \Phi_S \rangle\nonumber \\
&=&\prod_{j'=1}^M\frac{-i\gamma_T^*\gamma_Sg/v_T}{(i\nu_{j'} + \barGamma_T)(i\nu_{j'}+\barGamma_S)+|g|^2} \nonumber \\
&\times&\prod_{k'=M+1}^N\frac{i\gamma_T\gamma_S^*g^*/v_T}{(-i\nu_{k'} + \barGamma_T)(-i\nu_{k'}+\barGamma_S)+|g|^2} \nonumber \\
&\times&
\langle \Phi_S \vert \prod_{j=1}^{M}\hat{a}_S^\dagger(\nu_j)\prod_{k=M+1}^N\hat{a}_S(\nu_k)\vert\Phi_S
\rangle.
\end{eqnarray}
Provided $\vert \Phi_S \rangle$ describes a state containing photons at frequencies well within one linewidth of $\omega_S$, we can set $\nu_{j'}=\nu_{k'}=0$ in the denominators of the first two products in this expression, giving
\begin{eqnarray}
\lefteqn{\langle\mathcal{O}(\nu_1,...,\nu_N)\rangle}\nonumber \\
&=&\frac{(-i\gamma_T^*\gamma_Sg/v_T)^M(i\gamma_T\gamma_S^*g^*/v_T)^{N-M}}{(\barGamma_T\barGamma_S+|g|^2)^{N}} \nonumber \\
&\times& \langle \Phi_S \vert \prod_{j=1}^{M}\hat{a}_S^\dagger(\nu_j)\prod_{k=M+1}^N\hat{a}_S(\nu_k)\vert\Phi_S
\rangle,
\end{eqnarray}
which is, up to a frequency-independent proportionality factor, precisely the same function of $(\nu_1,...\nu_N)$ that the expectation value of $\langle\mathcal{O}(\nu_1,...,\nu_N)\rangle$ would be, were it calculated using the source operators $\hat{a}_S$ instead of $\hat{c}_T$. Provided the system is strongly over-coupled so that $\barGamma_J\approx\Gamma_J$ for $J=S,T$, making loss negligible, and the group velocities near the source and target frequencies are equal ($v_S\approx v_T$), the proportionality factor has unit magnitude when $|g|=\sqrt{\barGamma_S\barGamma_T}$. In such a system all measurable quantities relating to the input state of the source are effectively ``transplanted" into the target field: the device converts arbitrary inputs to the target frequency, not merely single photon states. However, for definiteness, in what follows we return to the case of a single photon input state.

\subsection{Spectral conversion probability}\label{subsec:gain_spectrum}
\begin{figure}
\includegraphics[width=1.0\columnwidth]{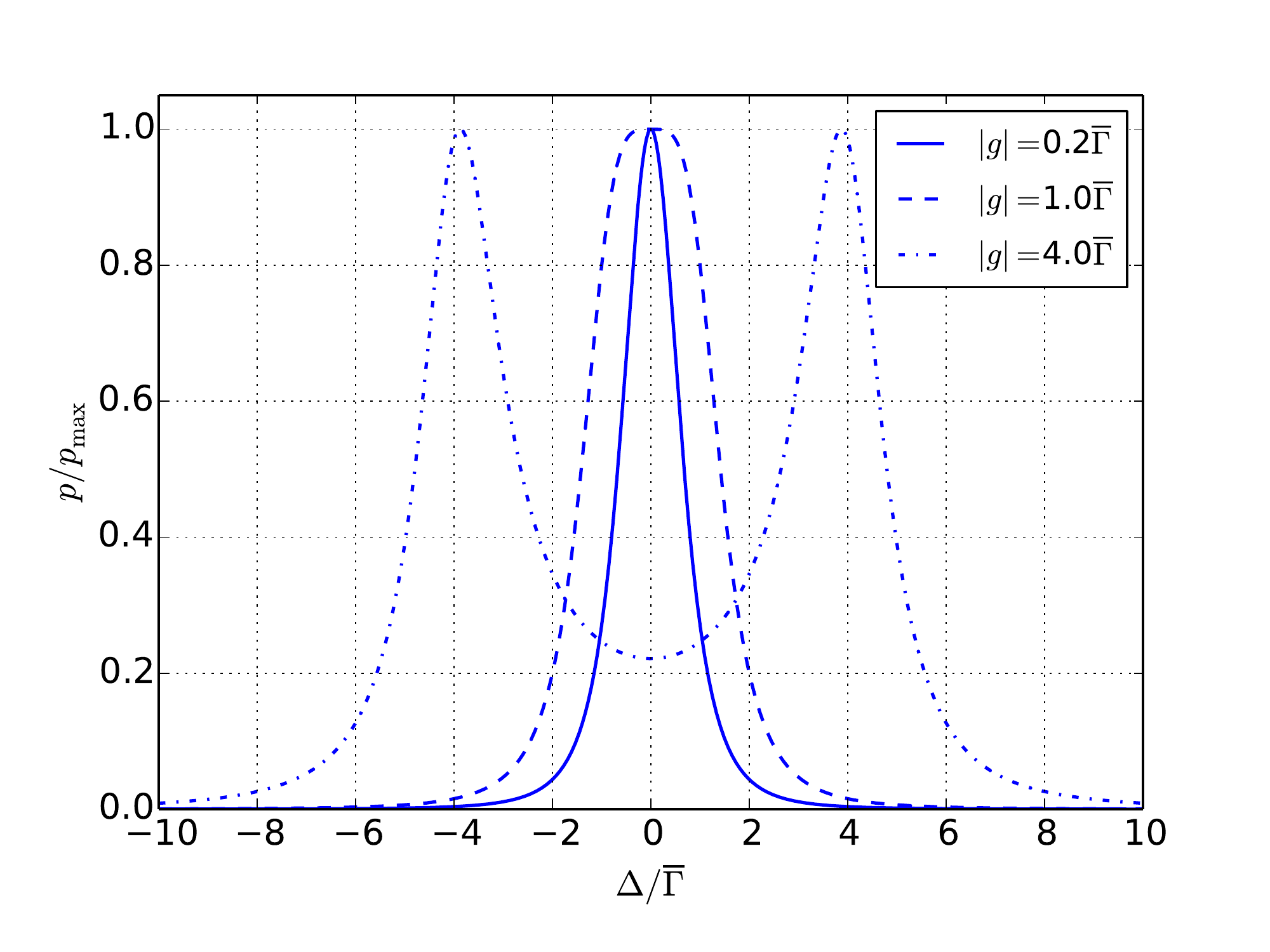}\caption{Device spectral sensitivity function $p(\Delta)$ (\ref{eqn:device_spec_defn}) scaled to unit maximum, plotted versus $\Delta$ in units of $\barGamma$ (taken to be equal for the source and target modes, $\barGamma_S=\barGamma_T\equiv\barGamma$) for different nonlinear coupling strengths $|g|$. Below a critical strength $|g|=\barGamma$ the spectrum is singly peaked at a frequency corresponding to the source mode resonance $\omega_S$ (solid curve). Above this critical $|g|$, when the source and target modes are strongly coupled, the spectrum is doubly peaked at frequencies approximately corresponding to $\omega_S\pm|g|$ (dot-dashed curve). When $|g|=\barGamma$ the spectrum exhibits a single, flat-topped peak that is significantly broadened (dashed curve), indicating a range of insensitivity to source photon input frequency.}\label{fig:conv_spectrum}
\end{figure}

Since the microring system has a finite bandwidth, responding only to inputs in a narrow frequency range about the source resonance, it is instructive to study the form of the \emph{spectral conversion probability density} $C(\Delta)$, which we take to be the expectation value of outgoing target photon number density as a function of frequency:
\begin{eqnarray}\label{eqn:conversion_spectrum_defn}
C(\Delta)=v_T\langle \hat{c}_T^\dagger(\Delta) \hat{c}_T(\Delta) \rangle.
\end{eqnarray}
Computing this for a single photon source input state $\vert \Phi_S \rangle$ with spectral profile $\hat{f_0}(\Delta)$,
\begin{eqnarray}\label{eqn:input_state}
\vert \Phi_S \rangle = \sqrt{v_S}\int \hat{f_0}(\Delta)\hat{a}_S^\dagger(\Delta) \vert \mathrm{vac}\rangle,
\end{eqnarray}
in which $\hat{f_0}(\Delta)$ is normalized according to $\int d\Delta |\hat{f}_0(\Delta)|^2=1$, we obtain
\begin{eqnarray}\label{eqn:conversion_spectrum_result}
C(\Delta)= 4\Gamma_S\Gamma_T|\hat{f}_0(\Delta)|^2p(\Delta),
\end{eqnarray}
where
\begin{eqnarray}\label{eqn:device_spec_defn}
p(\Delta)=\frac{|g|^2}{\big\vert(-i\Delta+\barGamma_T)(-i\Delta+\barGamma_S)+|g|^2\big\vert^2}.
\end{eqnarray}
The spectral conversion probability density is thus proportional to the product of the power spectrum of the input source photon with the factor $p(\Delta)$,
which is independent of the source photon input and describes the sensitivity of the response of the device as a function of the input frequency. As exhibited in Fig. \ref{fig:conv_spectrum}, plotting $p(\Delta)$ with $\barGamma_S=\barGamma_I\equiv\barGamma$ for different values of the source-target coupling strength $|g|$ reveals that the spectral response of the device is singly peaked when $|g|$ is smaller than the damping rate $\barGamma$, splitting into two peaks separated by approximately $2|g|$ when $|g|$ exceeds $\barGamma$, indicating the strong coupling of the source and target modes. At a  critical driving strength when $|g|=\barGamma$, $p(\Delta)$ exhibits a broad, flat-topped peak, indicating that the spectral response is nearly frequency-independent in a significant spectral range about resonance; the FWHM of this peak is $2\sqrt{2}\;\barGamma$, which is $\sqrt{2}$ times that of the ring resonance itself. The first three, and the fifth through seventh derivatives of $p(\Delta)$ at $\Delta=0$ are all precisely zero, indicating that the device spectrum (a smooth function) is extremely insensitive to frequency near $\Delta=0$ at this special $|g|$. As we demonstrate in the following section, this critical value of $|g|$ that gives rise to the flat spectral response is precisely the coupling strength that maximizes the probability of successfully converting the source photon to the target mode.

\subsection{Conversion probability}\label{subsec:conversion_probability}
\begin{figure}
\includegraphics[width=1.0\columnwidth]{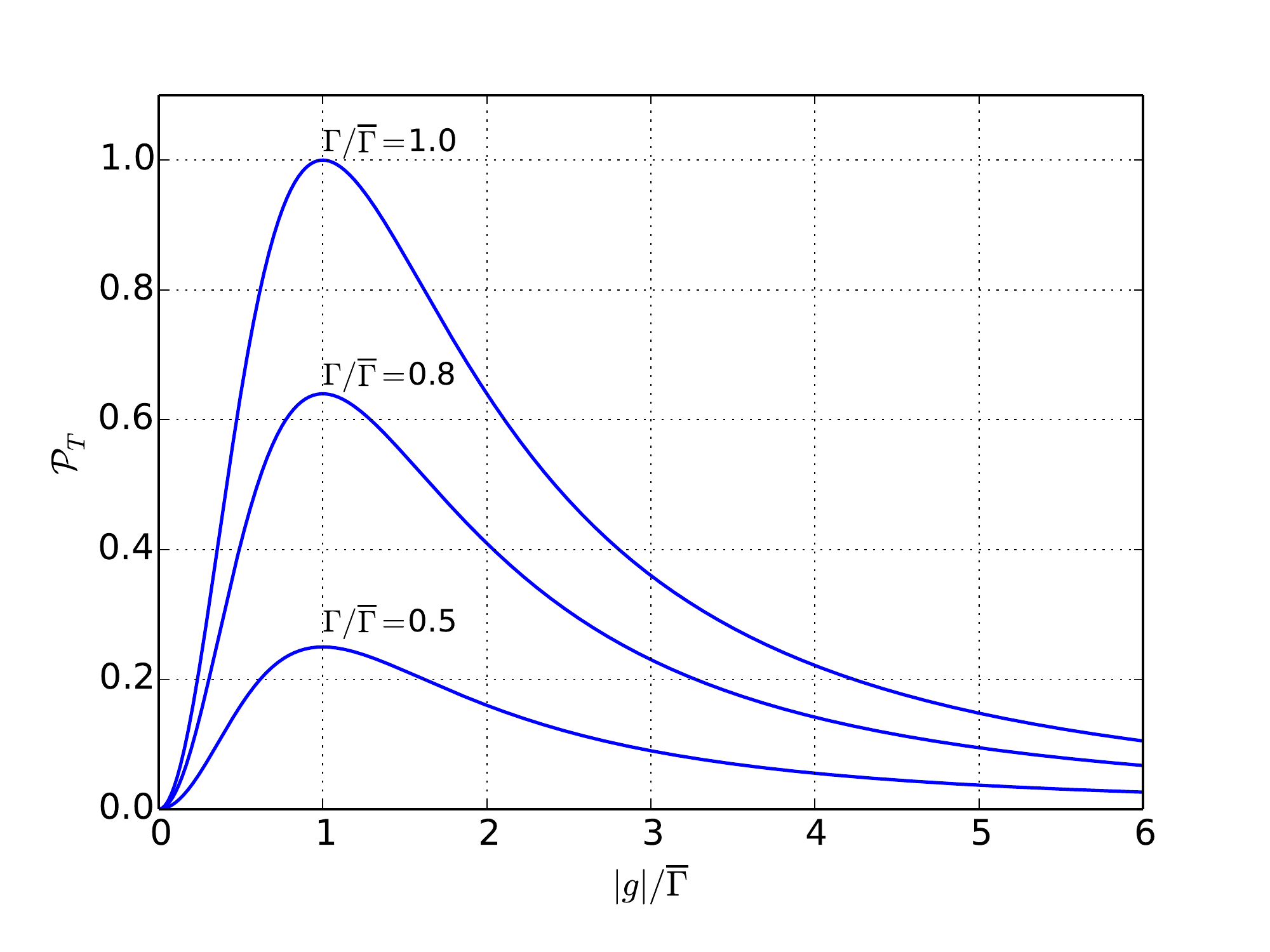}
\caption{Probability of successful conversion $\mathcal{P}_T$ (\ref{eqn:P_T_result_general}) as a function of coupling strength $|g|$ for devices with different coupling specifications. The coupling rates for the source and target were taken to be equal ($\Gamma_S=\Gamma_T\equiv\Gamma$ and $\barGamma_S=\barGamma_T\equiv\barGamma$). The maximum achievable conversion probability is limited only by loss, and tends to unity as the channel-ring coupling $\Gamma$ is increased. }\label{fig:conv_prob}
\end{figure}

Having calculated $C(\Delta)$, it is a straightforward matter to obtain the probability $\mathcal{P}_T$ of successful conversion of a source photon to the target mode. This is simply
\begin{eqnarray}\label{eqn:P_T_integral}
\mathcal{P}_T=\int C(\Delta) d\Delta.
\end{eqnarray}
Provided the spectrum $|\hat{f}_0(\Delta)|^2$ of the incoming photon is centred on frequency $\Delta_\mathrm{in}$ (corresponding to a source photon with central frequency $\omega_S + \Delta_\mathrm{in}$) and significantly narrower than the width of the device spectral response function $p(\Delta)$, we obtain
\begin{eqnarray}\label{eqn:P_T_result_general}
\mathcal{P}_T = \frac{4\Gamma_S\Gamma_T|g|^2}{\big\vert (-i\Delta_\mathrm{in} + \barGamma_T)(-i\Delta_\mathrm{in} + \barGamma_S) + |g|^2\big\vert^2}.
\end{eqnarray}
When the source photon central frequency is exactly $\omega_S$, i.e. $\Delta_\mathrm{in}=0$, this becomes
\begin{eqnarray}\label{eqn:P_T_defn}
\mathcal{P}_T = \frac{4\Gamma_S\Gamma_T|g|^2}{(\barGamma_S\barGamma_T + |g|^2)^2}.
\end{eqnarray}
This probability is plotted as a function of $|g|$ in Fig. \ref{fig:conv_prob}, and is maximum when the coupling strength reaches the critical value of $|g|=\sqrt{\barGamma_S\barGamma_T}$; this maximum is precisely
\begin{eqnarray}\label{eqn:P_T_max}
\mathcal{P}_T^{\mathrm{max}} = \frac{\Gamma_S\Gamma_T}{\barGamma_S\barGamma_T} = \frac{Q_SQ_T}{Q_S^\mathrm{ext}Q_T^\mathrm{ext}}.
\end{eqnarray}
The maximum achievable success probability is limited only by the ratio between the scattering and ring-channel coupling rates: When the ring system is strongly over-coupled with $\barGamma_S\approx \Gamma_S$ and $\barGamma_T\approx\Gamma_T$, so that $Q_S \approx Q_S^\mathrm{ext}$ and $Q_T \approx Q_T^\mathrm{ext}$, the maximum success probability is unity. The input power necessary to achieve this maximum conversion probability can be found by relating the coupling strength $|g|$ that maximizes $\mathcal{P}_T$ to $P^\mathrm{in}_{\pone}$ and $P^\mathrm{in}_{\ptwo}$; we find that the maximum conversion probability occurs when
\begin{eqnarray}\label{eqn:P_product_required}
P^\mathrm{in}_{\pone}P^\mathrm{in}_{\ptwo}=(\hbar\omega_\pone )(\hbar\omega_\ptwo )\frac{\barGamma_S\barGamma_T\barGamma_\pone ^2\barGamma_\ptwo ^2}{4\Lambda^2\Gamma_\pone \Gamma_\ptwo }.
\end{eqnarray}

Obtaining near-unit conversion probability in a realistic microring system requires a number of conditions to be satisfied. An ideal system would be strongly overcoupled for the source and target modes, but critically coupled for the pumped modes ($\barGamma_\pone =2\Gamma_\pone $ and $\barGamma_\ptwo =2\Gamma_\ptwo $). This is difficult to achieve in practice, for typical devices are usually fabricated to obtain the desired channel-ring coupling at one resonance. As demonstrated by Li \emph{et al.} \cite{Li2016}, it is possible to use a pulley-coupling scheme to achieve fairly similar quality factors across a wide spectral range while maintaining the overcoupling of the source and target modes. Yet this comes at the cost of overcoupling the pumped modes, manifesting the trade-off between power efficiency and conversion probability: as described by Eq. (\ref{eqn:P_product_required}), increasing the channel-ring coupling  necessitates higher input power required to reach the coupling strength that yields maximum conversion probability. This trade-off is analogous to that which arises between the heralding efficiency and heralding rate in microresonator-based heralded single photon sources \cite{Vernon2016,Steidle2016,He2015,Reimer2014,Savanier2016}. 

Despite these limitations, Li \emph{et al.} \cite{Li2016} have reported a conversion efficiency of over 60\% from a weak coherent source beam at 1550 nm to a target mode at 980 nm in silicon nitride microrings using less than 60 mW of input pump power, validating such systems as very promising candidates for integrated QFC. Comparison of those experimental results with estimates made using the calculations developed in this paper shows good agreement between theory and experiment (see Table \ref{tbl:specs}). While the efficiency of 60\% reported by Li \emph{et al.} slightly exceeds our calculated maximum of 49\% based on their reported quality factors, the inherent uncertainty involved in experimentally determining the system parameters, especially the insertion losses, leaves a fairly wide margin for error. With this in mind, we also explored other possible parameter values: excitingly, the parameters needed to obtain a success probability exceeding 95\% with less than 100 mW of input power are not far from the current state of the art. 

\subsection{Phase stability}\label{subsec:phase_stability}
While the conversion probability depends on the modulus of the coupling strength $g$, it is important to note that the phase of $g$ is relevant for certain applications. This phase, which is completely determined by the phases of the pump fields in the ring, must be stable on a timescale greater than the lifetimes $\barGamma_J^{-1}$ of the ring modes for conversion to efficiently take place. While this phase can be adjusted by suitably modulating the input beams, in practice it is difficult to achieve a definite phase relationship between two pumps produced by separate lasers. However, for the scheme considered in which the target frequency is approximately double the source frequency, it is possible to use a frequency-doubled version of $P^{(1)}$ for the second pump $P^{(2)}$, eliminating the need to actively control the each pump phase separately.

Still, even if a stable relative phase between the two pumps is achieved, the absolute phase stability of the pumps may be important. If it is crucial for a specific application to maintain the phase coherence between different photons injected and converted at different times in one experimental run, the pump phases must be stable over the entire duration between those times; an unstable pump would destroy any such phase coherence. 

\begin{table}
\begin{tabular}{|c|c|c|c|}
	\hline
	$Q_{S,\pone }^{\mathrm{int}}$ $(Q_{S,\pone })$ & $Q_{T,\ptwo }^{\mathrm{int}}$ $(Q_{T,\ptwo })$ & $P^{\mathrm{in}}$ (mW) & $\mathcal{P}_T^\mathrm{max}$ \\ \hline
	$\medmuskip=0mu 5\times 10^6$ ($\medmuskip=0mu 1\times 10^5$) & 	$\medmuskip=0mu 5\times 10^6$ ($\medmuskip=0mu 1\times 10^5$) & 65 & 0.96 \\ \hline
	$\medmuskip=0mu 3\times 10^6$ ($\medmuskip=0mu 3\times 10^5$) & $\medmuskip=0mu 1\times 10^6$ ($\medmuskip=0mu 1\times 10^5$) & 23 & 0.81 \\ \hline
	$\medmuskip=0mu 4.5\times 10^5$ ($\medmuskip=0mu 1.5\times 10^5$) & $\medmuskip=0mu 9.0\times 10^5$ ($\medmuskip=0mu 2.4\times 10^5$) & 28 & 0.49 \\ \hline
\end{tabular}
\caption{Table of approximate input power $P^\mathrm{in}=\sqrt{P_{\pone}^\mathrm{in}P_\ptwo^\mathrm{in}}$ required in the pump input fields to achieve maximum conversion probability $P_T^{\mathrm{max}}$ for several combinations of intrinsic (full) quality factors $Q_{S,\pone }^{\mathrm{int}}$ ($Q_{S,\pone }$) for the source and $\pone $ and $Q_{S,\ptwo }^{\mathrm{int}}$ ($Q_{S,\ptwo }$) for the target and $\ptwo$. Quality factors for nearby modes are assumed to be equal, and material parameters used correspond to those for typical silicon nitride microrings. The first row corresponds to idealized parameters, demonstrating that very high efficiency QFC at sub-watt input power is possible with only modest improvement to the current state of the art \cite{Gondarenko2009}. The last row corresponds to our estimates using parameters reported by Li \emph{et al.} \cite{Li2016}; these estimates for input power and conversion probabilities are in reasonbly good agreement with those reported experimental results.} \label{tbl:specs}
\end{table}
\section{Dressed modes}\label{sec:dressed_modes}
The simple frequency-domain approach used in the previous section is sufficient to calculate the conversion probability, which is the primary figure of merit for a microring QFC device. However, several qualitative aspects of the conversion dynamics, such as the dependence of $\mathcal{P_T}$ on the coupling strength $|g|$ shown in Fig. \ref{fig:conv_prob}, can be better understood using an alternate approach. For example, in contrast to conventional non-resonant QFC schemes, wherein the conversion probability oscillates as a function of pump power \cite{Clark2013}, the microresonator-based QFC system attains a single maximum conversion probability at one specific coupling strength, and declines asymptotically to zero as the input power exceeds this value. As suggested by Huang \emph{et al.} \cite{Huang2013,Huang2010,Huang2012}, this can be understood as a consequence of an effective shift in the resonance of ring modes that are coupled via the pumped modes. To fully explain this, here we develop a dressed mode picture for the conversion dynamics, identifying new modes in the ring which are linear combinations of the original source and target modes. These new modes are uncoupled, and represent new energy-shifted eigenmodes that couple to similar linear combinations of the source and target fields in the channel. The conversion dynamics, including the behavior of the conversion probability as function of coupling strength, can then be understood as a consequence of the phase shift imposed on incident source photons.

For the sake of clarity, in this section we develop our results for a system with equal coupling coefficients and group velocities for the source and target modes: $\barGamma_S=\barGamma_T\equiv\barGamma$, $\gamma_S=\gamma_T\equiv \gamma$, $\mu_S=\mu_T\equiv\mu$, and $v_S=v_T\equiv v$. The generalization to arbitrary coefficients is straightforward, and our conclusions do not depend sensitively on these assumptions.

The coupled system (\ref{eqn:S_T_ring_eqns}) can be written in matrix form,
\begin{eqnarray}\label{eqn:S_T_matrix_eqn}
\frac{d\mathbf{b}(t)}{dt}=\mathcal{Q}\mathbf{b}(t) + \mathbf{d}(t),
\end{eqnarray}
where $\mathbf{b}(t)=(\barb_S(t),\barb_T(t))^T$,
\begin{eqnarray}\label{eqn:Q_defn}
\mathcal{Q}=\begin{pmatrix}
-\barGamma && ig \\ ig^* && -\barGamma
\end{pmatrix},
\end{eqnarray}
and
\begin{eqnarray}\label{eqn:d_defn}
\mathbf{d}(t) = \begin{pmatrix}
-i\gamma^*\barpsi_{S<}(t) - i\mu^*\barphi_{S<}(t) \\
 -i\gamma^*\barpsi_{T<}(t) - i\mu^*\barphi_{T<}(t) \end{pmatrix}.
\end{eqnarray}
As is often done when solving coupled harmonic oscillator equations of motion \cite{Tikochinsky1979}, this system can be decoupled by diagonalizing $\mathcal{Q}$, resulting in the system of equations
\begin{eqnarray}\label{eqn:transformed_S_T_eqn}
\frac{d\mathbf{b}'(t)}{dt}=\mathcal{Q}'\mathbf{b}'(t) + \mathbf{d}'(t),
\end{eqnarray}
where 
\begin{eqnarray}\label{eqn:Q_prime}
\mathcal{Q}'=\begin{pmatrix}
-\barGamma + i|g| && 0 \\ 0 && -\barGamma - i|g|
\end{pmatrix}
\end{eqnarray}
and
\begin{eqnarray}\label{eqn:b_prime_defn}
\mathbf{b}'(t) = \frac{1}{\sqrt{2}}\begin{pmatrix}
e^{-i\theta}\barb_S(t) + \barb_T(t) \\
e^{-i\theta}\barb_S(t) - \barb_T(t)
\end{pmatrix} \equiv \begin{pmatrix}
\barb_+(t) \\ \barb_-(t)
\end{pmatrix}.
\end{eqnarray}
The transformed channel terms become
\begin{widetext}
\begin{eqnarray}
\mathbf{d}'(t) &=& \frac{1}{\sqrt{2}}\begin{pmatrix}
e^{-i\theta}(-i\gamma^*\barpsi_{S<}(t) - i\mu^*\barphi_{S<}(t)) +  (-i\gamma^*\barpsi_{T<}(t) - i\mu^*\barphi_{T<}(t)) \\
e^{-i\theta}(-i\gamma^*\barpsi_{S<}(t) - i\mu^*\barphi_{S<}(t)) -  (-i\gamma^*\barpsi_{T<}(t) - i\mu^*\barphi_{T<}(t))
\end{pmatrix} \nonumber \\ 
&\equiv& \begin{pmatrix}
-i\gamma^*\barpsi_{+<}(t) - i\mu^*\barphi_{+<}(t) \\
-i\gamma^*\barpsi_{-<}(t) - i\mu^*\barphi_{-<}(t) \end{pmatrix},
\end{eqnarray}
\end{widetext}
wherein we have introduced new channel fields $\barpsi_{\pm<}(t)=2^{-1/2}(e^{-i\theta}\barpsi_{S<}(t)\pm\barpsi_{T<}(t))$, and similar for $\barphi_{\pm<}(t)$. As pointed out in Sec. \ref{subsec:phase_stability},  the phase $e^{i\theta}=g/|g|$ that arises from the pump input beam phases is not especially relevant to our discussion; we thus assume the pump beams have been set such that $\theta=0$.

\begin{figure}
\includegraphics[width=1.0\columnwidth]{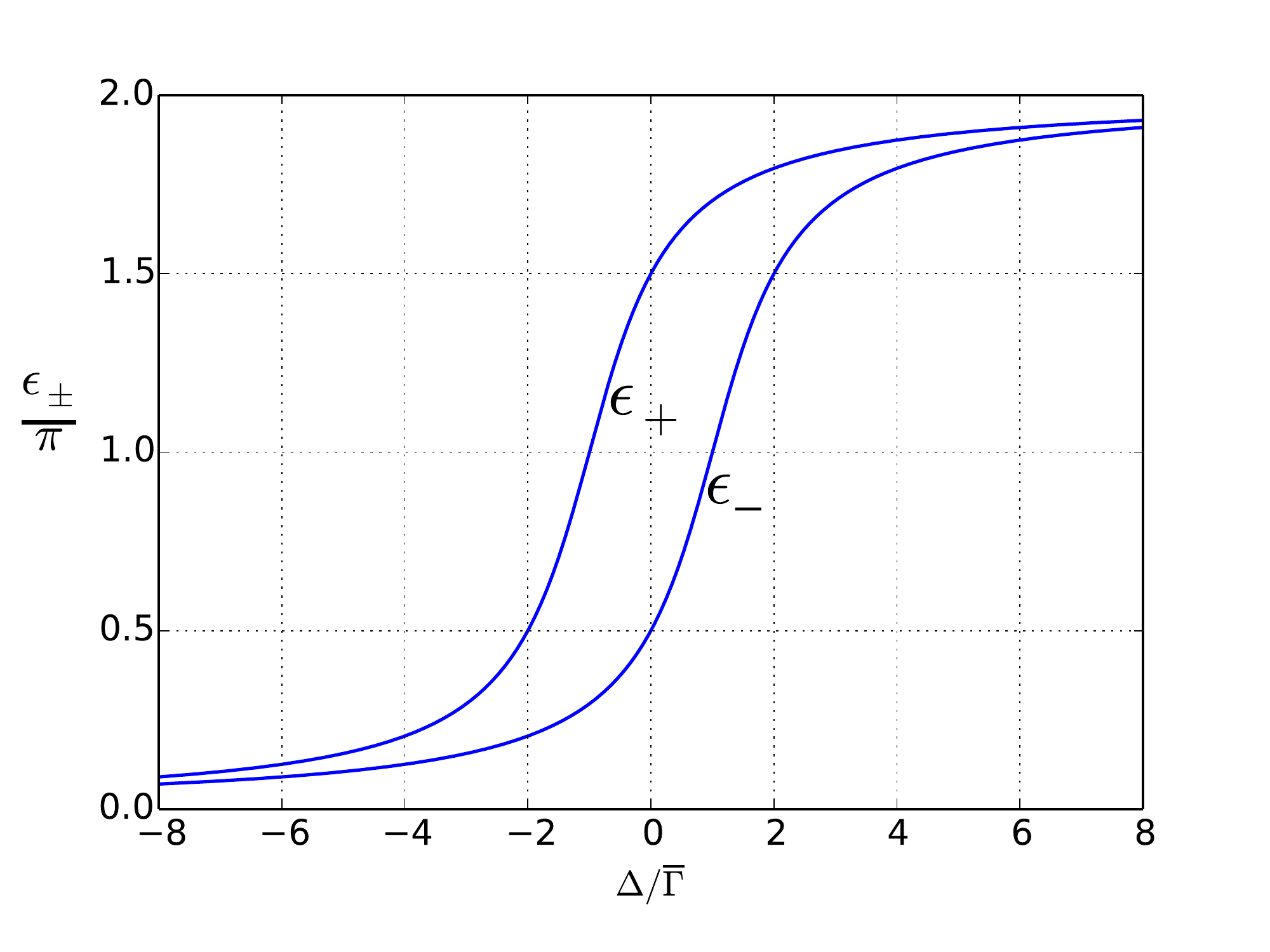}
\caption{Phase shift $\epsilon_\pm$ (\ref{eqn:phase_shifts}) imposed on incident photons in the $\barpsi_\pm$ field as a function of frequency for coupling strength $|g|=\barGamma$. At $\Delta=0$, corresponding to an incident source photon with frequency $\omega_S$, the total phase shift $\epsilon_+-\epsilon_-=\pi$, yielding complete frequency conversion to the target.}\label{fig:phase_shift}
\end{figure}

The diagonalized source-target mode system gives rise to new modes $\barb_\pm(t)$, which can be understood as equal symmetric and antisymmetric superpositions of the original source and target modes. These modes are shifted in energy by $\mp \hbar|g|$ from the original modes, and couple to similar equal symmetric and antisymmetric superpositions of the source and target channel fields as described by $\barpsi_{\pm<}(t)$. The form of these new channel fields is a feature inherited from the dressed modes in the ring, which determines the most natural combination of the channel fields to be used in writing the dressed mode dynamics. The response of the ring system to photons in these $\barpsi_{\pm<}$ fields incident from the side channel can be understood in the usual way one analyses a passive linear microring filter. Rewriting the channel input-output relation (\ref{eqn:channel_transformation}) in terms of the $\barpsi_{\pm<}$ fields, we obtain
\begin{eqnarray}\label{eqn:dressed_channel_transformation}
\barpsi_{\pm>}(t) = \barpsi_{\pm<}(t) -\frac{i\gamma}{v}\barb_\pm(t),
\end{eqnarray}
which becomes, in the frequency domain,
\begin{eqnarray}\label{eqn:dressed_channel_transformation_freqdomain}
\hat{c}_\pm(\Delta) = \hat{a}_\pm(\Delta) - \frac{i\gamma}{v}\hat{b}_\pm(\Delta),
\end{eqnarray}
where $\hat{c}_\pm(\Delta)$ and $\hat{a}_\pm(\Delta)$ are respectively the annihilation operators associated with the outgoing and incoming $\barpsi_\pm$ channel fields. These operators do not correspond to channel modes with a definite optical frequency offset of $\Delta$ from any single ring resonance, but rather to equal superpositions of channel modes offset by $\Delta$ from the source and target frequencies. Solving the transformed system of equations (\ref{eqn:transformed_S_T_eqn}) in the frequency domain, and substituting the resultant amplitudes $\hat{b}_\pm(\Delta)$ into (\ref{eqn:dressed_channel_transformation_freqdomain}), we obtain
\begin{eqnarray}\label{eqn:dressed_output_channel}
\hat{c}_\pm(\Delta)&=&\left(1-\frac{2\Gamma}{-i(\Delta \pm |g|) + \barGamma}\right)\hat{a}_\pm(\Delta) \nonumber \\
&-& \left(\frac{\mu^*\gamma}{v}\frac{1}{-i(\Delta \pm |g|) + \barGamma}\right)\hat{d}_\pm(\Delta).
\end{eqnarray}
For a strongly over-coupled system with $\Gamma \gg M$ so that $\barGamma\approx\Gamma$ (as is required to achieve high efficiency QFC), we can neglect the contribution from loss, giving
\begin{eqnarray}
\hat{c}_\pm(\Delta)=\frac{-i(\Delta\pm|g|) - \barGamma}{-i(\Delta\pm |g|)+\barGamma}\hat{a}_\pm(\Delta).
\end{eqnarray}
The factor multiplying $\hat{a}_\pm(\Delta)$ in this expression has unit modulus, serving only to impose a frequency-dependent phase on an incident photon in the $\barpsi_{\pm<}$ field that passes the ring to the outgoing $\barpsi_{\pm>}$ field. This phase shift is plotted in Fig. \ref{fig:phase_shift} as a function of frequency, and ranges between $0$ and $2\pi$ over a frequency range determined by $\barGamma$ centred on $\Delta=\mp |g|$, at which the phase shift is precisely $\pi$.

The input state $\vert \Phi_S\rangle$ (\ref{eqn:input_state}), which represents a photon with support in the frequency domain only near the source frequency, can be written in terms of the new channel fields, 
\begin{eqnarray}\label{eqn:input_state_transformed}
\vert \Phi_S\rangle = \sqrt{v}\int d\Delta \hat{f}_0(\Delta)\frac{1}{\sqrt{2}}(\hat{a}_+^\dagger(\Delta) + \hat{a}_-^\dagger(\Delta))\vert\mathrm{vac}\rangle.
\end{eqnarray}
After interacting with the microring, this state is transformed to the output state $\vert \Phi_T \rangle$
\begin{eqnarray}\label{eqn:output_state}
\lefteqn{\vert \Phi_T \rangle =} \\
& &\sqrt{v}\int d\Delta \hat{f}_0(\Delta)\frac{1}{\sqrt{2}}(\hat{a}_+^\dagger(\Delta)e^{i\epsilon_+} + \hat{a}_-^\dagger(\Delta)e^{i\epsilon_-})\vert\mathrm{vac}\rangle, \nonumber
\end{eqnarray}
where
\begin{eqnarray}\label{eqn:phase_shifts}
e^{i\epsilon_\pm}=\frac{-i(\Delta\pm|g|) - \barGamma}{-i(\Delta\pm |g|)+\barGamma}.
\end{eqnarray}
As illustrated in Fig. \ref{fig:phase_shift}, when the coupling strength equals the cavity damping rate, $|g|=\barGamma$, and for an input photon spectral profile $\hat{f}_0(\Delta)$ much narrower in extent than $\barGamma$, we have $\epsilon_+=3\pi/2$ and $\epsilon_-=\pi/2$, yielding an overall phase shift of $\pi$ between the $+$ and $-$ modes. The output state is then, up to an overall phase factor,
\begin{eqnarray}\label{eqn:output_state_reduced}
\vert \Phi_T\rangle &=& \sqrt{v}\int d\Delta \hat{f}_0(\Delta)\frac{1}{\sqrt{2}}(\hat{a}_+^\dagger(\Delta) - \hat{a}_-^\dagger(\Delta))\vert\mathrm{vac}\rangle \nonumber \\
&=& \sqrt{v}\int d\Delta \hat{f}_0(\Delta)\hat{a}_T^\dagger(\Delta)\vert\mathrm{vac}\rangle,
\end{eqnarray}
precisely the input state upconverted to the target mode.

The dressed mode pictures gives a clear explanation for the behaviour of the conversion probability as a function of the coupling strength $|g|$ shown in Fig. \ref{fig:conv_prob}. The phase shift incurred between the $+$ and $-$ components of an incoming source photon at $\omega_S$ is precisely $\pi$ only when $|g|=\barGamma$, at which point each of the $+$ and $-$ modes incur a $\pi/2$ phase shift. As $|g|$ increases past $\barGamma$, the frequencies of the dressed ring modes shift such that neither of them efficiently couples to the frequency range of the incoming source photon. The incoming photon is then entirely off resonance with all of the ring modes, and continues past the coupling point without the phase shift necessary to convert it to the target field. 

\section{Temporal dynamics}\label{sec:temporal_dynamics}
The temporal behaviour of the source and target modes in the ring can reveal interesting features of the conversion dynamics that are not immediately apparent in the frequency domain. Indeed, it is only in the time domain that an oscillatory regime is clearly demonstrated, in which the single photon input undergoes coherent oscillations between the source and target modes in a manner closely resembling Rabi oscillations \cite{Allen1987}. 

For clarity we again assume equal coupling constants, coupling rates and group velocities for different modes. The matrix equation of motion (\ref{eqn:S_T_matrix_eqn}) for the source and target mode operators can be solved exactly by introducing a Green functon $G(t,t')$, which takes the form of a $2\times 2$ matrix, such that the solution $\mathbf{b}(t)$ is
\begin{eqnarray}\label{eqn:green_solution}
\mathbf{b}(t)=\int_{-\infty}^t dt' G(t,t')\mathbf{d}(t').
\end{eqnarray}
The Green function must satisfy
\begin{eqnarray}
\frac{d}{dt}G(t,t')=\mathcal{Q}G(t,t')
\end{eqnarray}
for $t>t'$ subject to the initial condition $G(t,t')=I$, where $I$ is the $2\times 2$ identity matrix. Since the nonlinear coupling strength $|g|$ is constant for cw pumps, this equation has a simple solution,
\begin{eqnarray}\label{eqn:green_exponential}
G(t,t')=\exp\lbrace(t-t')\mathcal{Q}\rbrace,
\end{eqnarray}
which can be written explicitly as
\begin{eqnarray}\label{green_matrix}
\lefteqn{G(t,t')=} \\
& &e^{-(t-t')\barGamma}\begin{pmatrix}
\cos\left[(t-t')|g|\right] && ie^{i\theta}\sin\left[(t-t')|g|\right] \\
ie^{-i\theta}\sin\left[(t-t')|g|\right] && \cos\left[(t-t')|g|\right]
\end{pmatrix}.\nonumber
\end{eqnarray}
For simplicity, in the following we assume the pumps have been set such that the phase $e^{i\theta}=g/|g|=1$. The elements $G_{ij}$ of $G$ serve as temporal response functions that describe the evolution of the source and target modes as they couple to the channel fields and to each other. With an explicit expression for $G(t,t')$, the solutions for the source and target operators can be directly calculated using (\ref{eqn:green_solution}), giving
\begin{subequations}\label{eqn:S_T_time_evolution}
\begin{eqnarray}
\lefteqn{\barb_S(t) =} \\ 
& &\int_{-\infty}^t dt'\bigg[G_{11}(t,t')\left(-i\gamma_S^*\barpsi_{S<}(0,t') - i\mu_S^*\barphi_{S<}(0,t')\right) \nonumber \\
&+&\;\;\;\;\;\;\;G_{12}(t,t')\left(-i\gamma_T^*\barpsi_{T<}(0,t') - i\mu_T^*\barphi_{T<}(0,t')\right)\bigg], \nonumber \\
\lefteqn{\barb_T(t) =}\\ 
& &\int_{-\infty}^t dt'\bigg[G_{21}(t,t')\left(-i\gamma_S^*\barpsi_{S<}(0,t') - i\mu_S^*\barphi_{S<}(0,t')\right) \nonumber \\
&+&\;\;\;\;\;\;\;G_{22}(t,t')\left(-i\gamma_T^*\barpsi_{T<}(0,t') - i\mu_T^*\barphi_{T<}(0,t')\right)\bigg].\nonumber
\end{eqnarray}
\end{subequations}

\begin{figure}
\includegraphics[width=1.0\columnwidth]{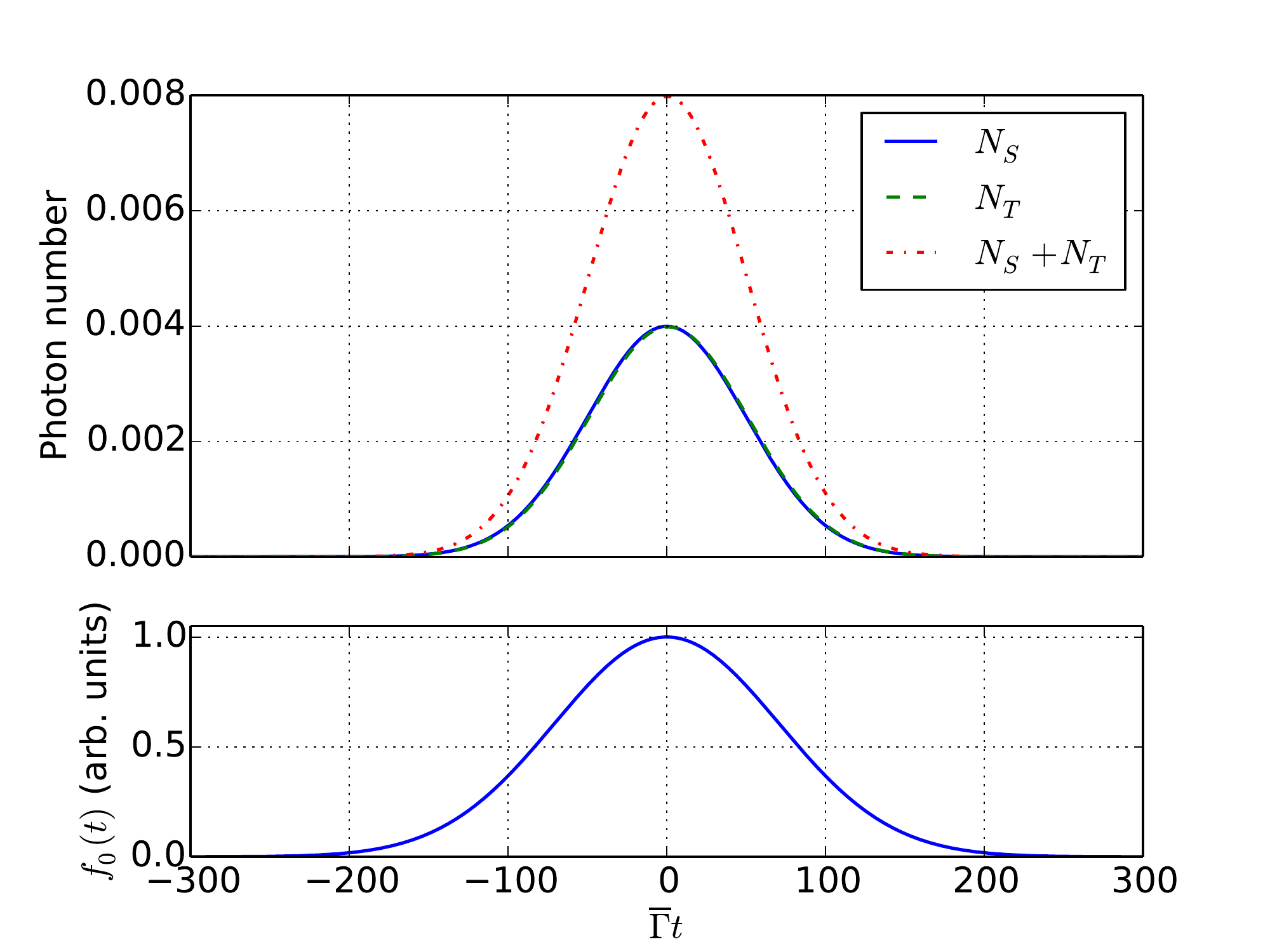}
\caption{Photon number expectation values $N_S(t)$ and $N_T(t)$ of source and target modes for input source photon with Gaussian temporal profile $f_0(t)\propto e^{-t^2/\tau^2}$ (plotted in lower panel) having long duration $\tau=100\barGamma^{-1}$. The coupling strength was taken to maximize the conversion probability, $|g|=\barGamma$, as discussed in Sec. \ref{subsec:conversion_probability}.}\label{fig:photon_num_maxprob}
\end{figure}

A single-photon source input state $\vert\Phi_S\rangle$ (\ref{eqn:input_state}) can be expressed in terms of its  temporal profile $f_0(t)$ as
\begin{eqnarray}\label{eqn:temporal_input_state}
\vert \Phi_S \rangle = \sqrt{v}\int dt' f_0(t') \barpsi_{S<}^\dagger(0,t')\vert \mathrm{vac}\rangle,
\end{eqnarray}
where $f_0(t)$ is normalized according to $\int dt' |f_0(t')|^2=1$. The photon number expectation values for the source and target modes can then be calculated explicitly for this state, giving
\begin{eqnarray}\label{eqn:N_S}
N_S(t) &=& \langle \barb_S^\dagger(t) \barb_S(t)\rangle \\
&=& 2\Gamma\left\vert\int_{-\infty}^t dt' e^{-(t-t')\barGamma}\cos\left[(t-t')|g|\right]f_0(t')\right\vert^2\nonumber
\end{eqnarray}
and
\begin{eqnarray}\label{N_T}
N_T(t) &=& \langle \barb_T^\dagger(t) \barb_T(t)\rangle \\
&=& 2\Gamma\left\vert\int_{-\infty}^t dt' e^{-(t-t')\barGamma}\sin\left[(t-t')|g|\right]f_0(t')\right\vert^2.\nonumber
\end{eqnarray}
The photon number is therefore calculated as the input photon temporal profile integrated against a response function that decays at the rate determined by the ring resonance linewidth, and oscillates at the coupling frequency $|g|$. Since at most one photon is ever present in the source and target modes, \emph{the functions $N_S(t)$ and $N_T(t)$ can be interpreted as the instantaneous probability at time $t$ for there to be a single photon in the source and target ring mode, respectively.} As illustrated in Fig. \ref{fig:photon_num_maxprob}, for an input source photon having a Gaussian temporal profile with duration greatly exceeding the ring mode lifetime $\barGamma^{-1}$ (and bandwidth much narrower than $\barGamma$), and with the coupling strength tuned to yield maximal conversion probability ($|g|=\barGamma$), $N_S(t)$ and $N_T(t)$ smoothly rise and fall as the source photon couples into the ring and is transferred to the target mode. 

\begin{figure}
\includegraphics[width=1.0\columnwidth]{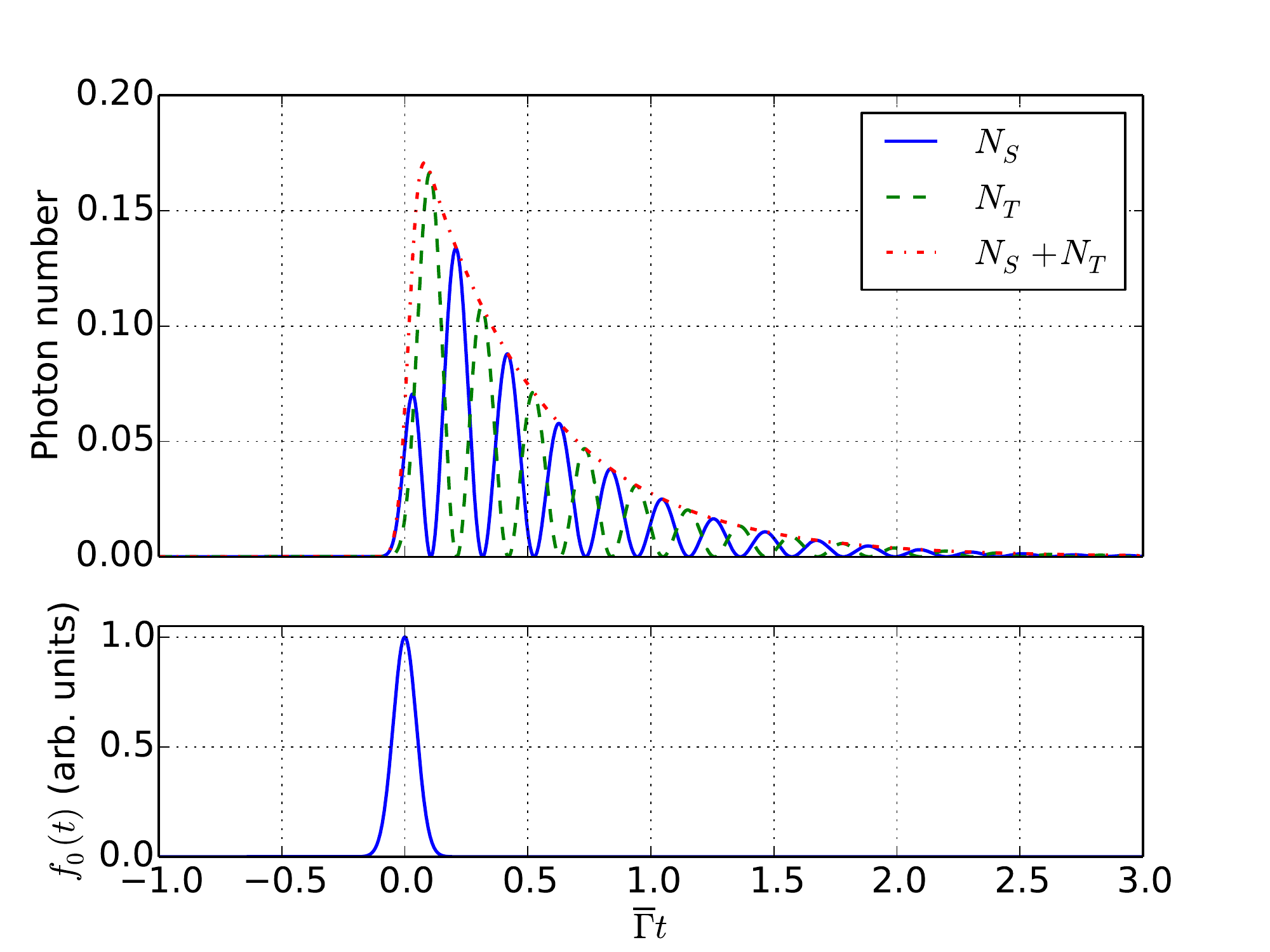}
\caption{Photon number expectation values $N_S(t)$ and $N_T(t)$ of source and target modes for input source photon with Gaussian temporal profile $f_0(t)\propto e^{-t^2/\tau^2}$ (plotted in lower panel) having short duration $\tau=0.1\barGamma^{-1}$. When the coupling strength $|g|$ is sufficiently large compared to the damping rate ($|g|=15\barGamma$ in this plot), the source and target photon numbers oscillate with frequency $|g|$ and are out of phase by $\pi$.}\label{fig:photon_num_oscillation}
\end{figure}

The behaviour of the photon probabilities in the ring is quite different for input source photons with shorter durations, and when the modes are more strongly coupled with $|g|>\barGamma$. As plotted in Fig. \ref{fig:photon_num_oscillation}, in this regime $N_S(t)$ and $N_T(t)$ oscillate out of phase by $\pi$ as the input photon is transferred back and forth between the source and target modes. This behaviour is strongly reminiscent of the Rabi oscillations that are displayed by two-level systems driven near resonance. Indeed, an analogy can be drawn between such systems and the QFC device: one can identify a ground state $\vert \mathrm{g}\rangle=b_S^\dagger\vert\mathrm{vac}\rangle$, and an excited state $\vert\mathrm{e}\rangle=b_T^\dagger\vert\mathrm{vac}\rangle$. These states have well-defined energies of $\hbar\omega_S$ and $\hbar\omega_T$, up to the precision permitted by the linewidths of the ring modes. Transitions between these states are driven by the pump beams, yielding Rabi oscillations at the frequency $|g|$. A similar perspective can be used to view these two optical states as comprising a qubit, in which 0 is represented by the presence of a photon in the source mode, and 1 by the presence of a photon in the target mode; such an approach has been taken by Clemmen \emph{et al.} in a nonresonant fibre-optic implementation \cite{Clemmen2016}. The QFC process can then be interpreted as implementing a Hadamard gate on the input qubit. We intend to study this oscillatory regime in more detail in the near future.

\section{Conclusion}\label{sec:conclusion}
We have studied the dynamics of quantum frequency conversion using four-wave mixing in microresonators, focusing especially on silicon nitride microrings. Three approaches were used: (i) a frequency-domain solution to the conversion dynamics enabled the calculation of the converison probability, spectral conversion probability density, and power requirements, (ii) a dressed mode formalism provided a clear intuitive explanation for the qualitative features of the conversion process, and (iii) a temporal analysis of the photon number expectation values revealed a regime of Rabi-like oscillations.

By suitably engineering the dispersion of the resonator and selecting appropriate pump frequencies and input powers, high efficiency wideband frequency translation of arbitrary quantum states with low noise was shown to be achievable with less than 100 mW of pump power; efficiencies exceeding 95\% using only 65 mW of power were predicted to be achievable with only modest improvement to the current state of the art. The maximum probability of successful conversion is limited only by loss, and is given by the product (\ref{eqn:P_T_max}) of the ratios between the extrinsic and full loaded quality factors of the resonator for the source and target modes; this maximum tends to unity as the microresonator-channel system is more strongly over-coupled. A simple expression (\ref{eqn:P_product_required}) for the required input power to achieve this maximum was derived, and the conversion probability as a function of coupling strength was shown to exhibit a single maximum followed by an asymptotic decay to zero for large input powers. These results are in good agreement with both previously developed theory \cite{Huang2013} and experiment \cite{Li2016}. The spectral conversion probability density that describes the conversion bandwidth of the device was calculated, and was found to exhibit a broad, flat-topped peak at the source mode frequency, indicating a spectral range where the device is very insensitive to  source input frequency. This enables efficient conversion of source photons even with complicated spectral profiles over a wider bandwidth than might naively be expected based on the unperturbed resonance linewidths.

The dressed mode picture was developed to better explain the qualitative features of QFC in microesonators. In this model the system of equations of motion for the source and target mode annihilation operators in the resonator was diagonalized, yielding new, uncoupled  and energy-shifted dressed modes that are linear combinations of the original modes. These couple to similar linear combinations of the channel fields, which obey an input-output relation formally identical to that of a passive linear microring filter. The frequency conversion process can then be understood as a consequence of the phase shift imposed between the different components of  the incoming source photon to be converted. The magnitude of this phase shift is dependent on the coupling strength $|g|$, and reaches the necessary value required for unit conversion probability only for $|g|=\barGamma$, where $\barGamma$ is the full damping rate of the resonator.

By directly studying the temporal evolution of the intraring photon number expectation values for the source and target modes, an oscillatory regime was revealed in which a single photon input oscillates between the two frequency modes at a rate determined by the coupling strength. This behaviour strongly resembles Rabi oscillations that are observed in a coherently driven two-level atom.

Fabrication techniques for microresonators are rapidly advancing, with new record quality factors, better dispersion engineering, and more extensive control over coupling conditions being routinely reported. With such progress we expect that microresonators will play an important role in future efforts to develop integrated quantum frequency conversion devices. 

\begin{acknowledgements}
This work was financially supported by the Natural Sciences and Engineering Research Council of Canada, and European COST Action MP 1403.
\end{acknowledgements}

\appendix
\section{Self- and cross-phase modulation}
In addition to the interaction (\ref{eqn:nonlinear_H}) that gives rise to desired QFC process, the full nonlinear Hamiltonian contains terms that correspond to self-phase modulation (SPM) of the pumped modes and cross-phase modulation (XPM) between the pumped modes and the source and target modes \cite{Vernon2015,Vernon2015b,Matsko2005,Chembo2010,Chembo2016}. The Hamiltonian describing SPM is given by \cite{HamiltonianNote}
\begin{eqnarray}\label{eqn:spm_hamiltonian}
\lefteqn{H_\mathrm{SPM}=} \\
& &-\hbar\eta_\ptwo  b_\ptwo  ^\dagger b_\ptwo  b_\ptwo ^\dagger b_\ptwo  - \hbar\eta_\pone  b_\pone ^\dagger b_\pone  b_\pone ^\dagger b_\pone ,\nonumber
\end{eqnarray}
where $\eta_\ptwo $ and $\eta_\pone $ are the coefficients associated with SPM.  Cross-phase modulation is described by
\begin{eqnarray}\label{eqn:XPM_hamiltonian}
\lefteqn{H_\mathrm{XPM}=} \\
&-&\hbar\zeta_{\ptwo S}b_\ptwo ^\dagger b_\ptwo  b_S^\dagger b_S - \hbar\zeta_{\ptwo T} b_\ptwo ^\dagger b_\ptwo  b_T^\dagger b_T \nonumber \\
&-&\hbar\zeta_{\pone S}b_\pone ^\dagger b_\pone  b_S^\dagger b_S - \hbar\zeta_{\pone T}b_\pone ^\dagger b_\pone  b_T^\dagger b_T \nonumber \\
&-&\hbar\zeta_{\ptwo \pone }b_\ptwo ^\dagger b_\ptwo  b_\pone ^\dagger b_\pone \nonumber,
\end{eqnarray}
where $\zeta_{JJ'}$ is the coefficient associated with XPM between modes $J$ and $J'$. In these expressions we have neglected terms that lead to SPM of the source and target modes, as well as XPM between those modes, since they never contain enough energy for these effects to be significant.

When the system is driven by cw pump beams, the effect of SPM and XPM is simply to shift the effective resonance frequencies of the ring modes by an amount determined by the number of photons present in those modes. The frequency shifts $\delta_J$ of the ring resonances are given by \cite{Vernon2015b}
\begin{eqnarray}
\delta_S&=&-\zeta_{\pone S}N_\pone -\zeta_{\ptwo S}N_\ptwo , \nonumber \\
\delta_T&=&-\zeta_{\pone T}N_\pone -\zeta_{\ptwo T}N_\ptwo , \nonumber \\
\delta_\pone &=&-\zeta_{\ptwo \pone }N_\ptwo -\eta_{\pone }N_\pone , \nonumber \\
\delta_\ptwo &=&-\zeta_{\ptwo \pone }N_\pone -\eta_{\ptwo }N_\ptwo ,
\end{eqnarray}
where $N_J$ is the steady-state photon number expectation value of mode $J$ in the ring; these are given by $N_J=|\barbeta_J|^2$, where $\barbeta_J$ is the amplitude of pumped ring mode $J$ (either $P^{(1)}$ or $P^{(2)}$) (\ref{eqn:pump_amplitudes}). By slowly tuning the frequency of the pump beams as their intensity is increased, the pumps can stay on resonance and continue to efficiently couple to the ring \cite{Vernon2015b}. The ring mode reference frequencies $\omega_J$ in this work can then be understood to include the effect of SPM and XPM. However, it is necessary to ensure that the energy-conservation relation (\ref{eqn:energy_conservation}) remains satisfied for the shifted resonances. We therefore require
\begin{eqnarray}\label{eqn:freq_shift_matching}
\delta_T - \delta_S = \delta_\ptwo  - \delta_\pone ,
\end{eqnarray}
which reduces to 
\begin{eqnarray}
& &(\zeta_{\pone S}-\zeta_{\pone T})N_\pone  + (\zeta_{\ptwo S}-\zeta_{\ptwo T})N_\ptwo  \\
&=&(\eta_\pone -\zeta_{\ptwo \pone })N_\pone  + (\zeta_{\ptwo \pone }-\eta_\ptwo )N_\ptwo .\nonumber
\end{eqnarray}
The XPM coeffiecient between the target and $\pone $ is very close to that between $\ptwo$ and $\pone $, since the target and $\ptwo$ are close in frequency, giving $\zeta_{\pone T}\approx\zeta_{\ptwo \pone }$; similarly, $\zeta_{\ptwo S}\approx\zeta_{\ptwo \pone }$. The relation (\ref{eqn:freq_shift_matching}) then becomes
\begin{eqnarray}\label{eqn:freq_shift_matching_reduced}
\lefteqn{\zeta_{\pone S}N_\pone  - \zeta_{\ptwo T}N_\ptwo} \\
& & \approx \eta_\pone N_\pone  - \eta_\ptwo N_\ptwo .\nonumber
\end{eqnarray}
To maintain energy conservation, we must therefore have
\begin{eqnarray}\label{eqn:photon_num_restriction}
\frac{N_\pone }{N_\ptwo }=\frac{\zeta_{\ptwo T}-\eta_\ptwo }{\zeta_{\pone S}-\eta_\pone }.
\end{eqnarray}
The ratio between the SPM and XPM coefficients for nearby modes is independent of frequency \cite{Andersen2015,Vernon2015b}, so this condition reduces to 
\begin{eqnarray}\label{eqn:reduced_photon_num_restriction}
\frac{N_\pone }{N_\ptwo }=\frac{\eta_\ptwo }{\eta_\pone }.
\end{eqnarray}
This can easily be achieved by adjusting the input power to the pump modes: using (\ref{eqn:pump_amplitudes}) we obtain for the required input power in the those modes
\begin{eqnarray}
\frac{P^\mathrm{in}_{\pone}}{P^\mathrm{in}_{\ptwo}}\approx 
\frac{\eta_\pone }{\eta_\ptwo }\frac{Q_\ptwo ^2Q_\pone ^\mathrm{ext}}{Q_\pone ^2Q_\ptwo ^\mathrm{ext}}.
\end{eqnarray}

\bibliography{FreqConversion}

\end{document}